\documentclass[aps,prd]{revtex4}
\usepackage{graphicx}
\usepackage{color}
\begin{document}

\title{Skewness as a test for Quartessence}
\author{R.R.R. Reis,  M. Makler and I. Waga}
\affiliation{Universidade Federal do Rio de Janeiro, \\
Instituto de F\'\i sica, \\
CEP 21941-972 Rio de Janeiro, RJ, Brazil}
\date{\today}

\begin{abstract}
Quartessence is one of the alternatives to $\Lambda$-CDM that has
lately attracted considerable interest. According to this unifying
dark matter/energy scenario, the Universe evolved from an early
non-relativistic matter-dominated phase to a more recent
accelerated expansion phase, driven by a single fluid component.
Recently, it has been shown that some problems of the quartessence
model, such as the existence of instabilities and oscillations in
the matter power spectrum, can be avoided if a specific type of
intrinsic entropy perturbation is considered. In the present
article we explore the role of skewness in constraining this
non-adiabatic scenario. We show that non-adiabatic quartessence
and quintessence have different signatures for the skewness of the
density distribution on large scales and suggest that this
quantity might prove helpful to break possible degeneracies
between them.
\end{abstract}
\maketitle

\section{Introduction}

According to the current standard cosmological model, the dynamics
of the universe would be dominated by two unknown components:
dark-matter (DM), responsible for structure formation, and
dark-energy (DE), that causes the accelerated expansion. Although
there are several candidates for both DM and DE, there is still no
evidence of either of them in laboratory physics. From the point
of view of simplicity, it would be interesting to explore the
possibility that a single component plays the role of both DE and
DM, reducing from two to one the unknown constituents of the
universe. A model that provides a single description of DE and DM
through \textquotedblleft unifying-dark-matter\textquotedblright\
or simply quartessence \cite{makler03} has attracted a lot of
interest recently. A prototype of this model is given by the
quartessence Chaplygin model (QCM) \cite{kamenshchik01}.

Both the background and linear fluctuations were extensively
studied for QCM, and were compared to observational data. The
generalized Chaplygin gas (as quartessence) appears to be
compatible with all available data regarding the expansion history
(see e.g. ref. \cite{makler03b} and refs. therein). For adiabatic
perturbations, a linear analysis was done for the CMB
\cite{amendola03}\ and LSS \cite{beca03}. In this case, only QCM
models close to the ``$\Lambda$CDM limit'' are allowed. Recently,
it was shown that problems (pointed out in \cite{sandvik02}), such
as the existence of instabilities and oscillations in the matter
power spectrum of QCM, can be avoided if a specific type of
intrinsic entropy perturbation is considered. Such non-adiabatic
model is consistent with the 2dF power spectrum for any value of
the model parameters in the permitted interval, as long as the
effective shape parameter assumes certain values \cite{reis03b}.
An \textquotedblleft averaging problem\textquotedblright\ was also
pointed out as a shortcoming of quartessence \cite{avelino03}.
However, it is straightforward to show that the above mentioned
non-adiabatic quartessence does not suffer this kind of problem
\cite{obs}.

Thus, up to the present time, we can say that adiabatic
quartessence is disfavored by the data, but it is not possible to
distinguish non-adiabatic quartessence from concordance models
like $\Lambda $CDM and quintessence using the previously
considered observables. However, as we shall see, measurable
differences in the predictions of these models appear clearly in
the nonlinear regime, in particular in the skewness of the matter
distribution in large scales. In our investigation we specifically
consider three different quartessence models. All of them have the
$\Lambda $CDM model as a limiting case for the background
solution. The analysis of these three cases indicates that our
result should be applicable to more generic quartessence models.

While most studies of the nonlinear regime deal with the clumping
of pressureless fluid (DM), in the case of quartessence it is
imperative that one includes the effects of pressure. Accordingly,
we apply, and somewhat extend to include relativistic pressure, a
method for the computation of density cumulants developed in refs.
\cite{bernardeau92,bernardeau94,fosalba98}.

\section{Gravitational growth for quartessence in the spherical collapse approximation}

In our approach we consider the following newtonian-like equations
\cite{lima,reis03}
\begin{equation}
\nabla_{r}^{2}\phi   =    4\pi G(\rho+3P)\;, \label{poisson}
\end{equation}
\begin{equation}
\left(\frac{\partial\vec{u}}{\partial t}\right)_{r} +
(\vec{u}\cdot\vec{\nabla}_{r})\vec{u}    = -\vec{\nabla}_{r}\phi -
(\rho + P)^{-1}\vec{\nabla}_{r}P\;, \label{euler}
\end{equation}
\begin{equation}
\left(\frac{\partial\rho}{\partial t}\right)_{r} +
\vec{\nabla}_{r}\cdot(\rho\vec{u}) + P\vec{\nabla}_{r}\cdot\vec{u}
= 0 \;.\label{cont}
\end{equation}
Equations (\ref{poisson}), (\ref{euler}) and (\ref{cont}) are,
respectively, the Poisson,  Euler and energy conservation
equations, where relativistic effects of pressure (inertia and
active gravitational mass) have been included. In these equations
$\rho$, $P$, $\vec{u}$ and $\phi$ stand for, the energy density,
pressure, velocity field, and gravitational potential of the
cosmic fluid, and we take $c=1$. In the linear regime, for
$c_{eff}=0$ (see bellow), system (\ref{poisson}-\ref{cont}) gives
exactly the same equations as the relativistic perturbation theory
in a particular gauge \cite{reis03}. Also the (nonlinear) energy
conservation and the Raychadhuri equations derived from this
system are formally identical to the general relativistic ones
\cite{ellis} in this case, which provides a motivation for the
above system.

It is useful to split the dynamical variables into their
background and inhomogeneus parts, i.e. we write: $ \rho =
\bar{\rho} + \delta\rho = \bar{\rho}(1+\delta)$, $P    = \bar{P} +
\delta P$, $\phi = \bar{\phi} + \varphi$, and $\vec{u}  =  H \,
\vec{r} + \vec{v}$. Here $H=a^{-1} \,da/dt$ is the Hubble
parameter, $a$ stands for the scale factor, and the overbar
``$\,\,\bar{}\,\,$'' denotes background (average) quantities.
Introducing comoving coordinates $\vec{x}=\vec{r}/a$, neglecting
shear and vorticity, and taking into account the background
equations (assuming critical density), we obtain, after some
algebra, the following differential equation for the density
contrast $\delta$,

\begin{eqnarray}
&&\left. \delta '' +\delta ' \left[ \frac{1- 9 w }{2}+ 3 c_{eff}^2
- \frac{w '}{1+w+\delta(1+c_{eff}^2)}\right]- {\delta '}^2 \left[
\frac{4/3+c_{eff}^2}{1+w+\delta(1+c_{eff}^2)}\right]\nonumber
\right.
\\ && \left. + \delta
\delta' \left[
\frac{(w-c_{eff}^2)(5+3c_{eff}^2)-{c_{eff}^2}'}{1+w+\delta(1+c_{eff}^2)}\right]+
\delta ^2 \left[ \frac{3(w-c_{eff}^2){c_{eff}^2}
'-3(w-c_{eff}^2)^2}{1+w+\delta(1+c_{eff}^2)}-\frac32(1+3c_{eff}^2)(1+c_{eff}^2)\right]\nonumber
\right. \\ &&\left. + \frac32\,\delta \left[
(3w^2-2w-1)-2c_{eff}^2(1+3w) -2(w'-{c_{eff}^2}')
+\frac{2(w-c_{eff}^2)w'}{1+w+\delta(1+c_{eff}^2)}\right]
\nonumber \right. \\
&&\left. =
\frac{1+w+\delta(1+c_{eff}^2)}{{\cal{H}}^2}\;\overrightarrow{\nabla}_x
\cdot\left(\frac{ \overrightarrow{\nabla}_x
\left(c_{eff}^2\,\delta\right)} {1+w+\delta(1+c_{eff}^2)}
\right).\right. \label{delta2}
\end{eqnarray}
In the above equation the prime symbol denotes differentiation
with respect to $\eta =\ln\, (a)$, ${\cal{H}} = a H$, $w=\bar P /
\bar \rho$ and $c_{eff}^2=\delta P / \delta \rho$ \cite{hu}. In
the linear approximation, in the special case in which
perturbations are adiabatic we have $c_{eff}^2=c_{s}^2\equiv
\bar{P}'/\bar{\rho}'=w-w'/3(1+w)$. However, as discussed in
\cite{sandvik02}, in this case, the right hand side of
(\ref{delta2}) gives rise to oscillations and instabilities in the
mass power spectrum that render the model unacceptable. This
problem can be circumvented if non-adiabatic perturbations such
that $c_{eff}^2=0$, are considered \cite{reis03b}. In the
following we assume this is the case such that Eq. (\ref{delta2})
simplifies to,

\begin{eqnarray}
&&\left. \delta '' +\delta ' \left[ \frac{1- 9 w }{2}- \frac{w
'}{1+w+\delta}\right]- {\delta '}^2 \left[
\frac{4/3}{1+w+\delta}\right]+ \delta \delta' \left[
\frac{5w}{1+w+\delta}\right]\nonumber \right.
\\ && \left. - 3\delta
^2 \left[\frac{w^2}{1+w+\delta}+\frac12\right]+ \frac32\,\delta
\left[ 3w^2-2w-1 -2w' +\frac{2w w'}{1+w+\delta}\right]=0 \right..
\label{delta2na}
\end{eqnarray}

To study the weakly nonlinear regime of structure formation and
compute the higher order moments of the density distribution, it
is useful to expand $\delta$ as
\cite{bernardeau92,bernardeau94,fosalba98},
\begin{equation}
\label{expansion}
\delta=\sum_{i=1}^{\infty}\delta_i=\sum_{i=1}^{\infty}\frac{D_i(\eta)}
{i!}\delta_{0}^{i}\;,
\end{equation}
where $\delta_{0}$ is a small perturbation. Using the above
expansion we obtain for the linear ($D_1$) and second order
($D_2$) factors the following differential equations,
\begin{equation}
\label{linear} D_1''+D_1'\left[ \frac{1- 9 w }{2}- \frac{w
'}{1+w}\right]+ \frac32\,D_1 \left[ 3w^2-2w-1 -\frac{2
w'}{1+w}\right]=0\;,
\end{equation}
and
\begin{eqnarray}
&&\left. {D_2}'' + {D_2}' \left[ \frac{1- 9 w }{2}- \frac{w
'}{1+w}\right]+  \frac32\,D_2 \left[ 3w^2-2w-1 -\frac{2
w'}{1+w}\right]- {D_1}'^2 \left[ \frac{8/3}{1+w}\right]\nonumber
\right.
\\ && \left. + D_1
{D_1}' \left[ \frac{10 w}{1+w}+\frac{2 w'}{(1+w)^2}\right]- 3{D_1}
^2 \left[\frac{2 w^2}{1+w}+\frac{2 w w'}{(1+w)^2}+1\right]=0
\right. .
\end{eqnarray}
Analogously, higher order modes can be obtained recursively by
using the solutions of the differential equations for the lower
order terms. Of special interest is the second-order equation. If
we start with Gaussian initial conditions, it is associated with
the emergence of non-Gaussian features in the matter density
field. Further, $D_2$ can be related to the skewness of the cosmic
field \cite{peebles,fry}. In this case, the unsmoothed skewness is
given by $ S_3=3D_2/D^{2}_{1}$.

\begin{figure}\centering \hspace*{-0.8in}
\includegraphics[height= 8 cm,width=11cm]{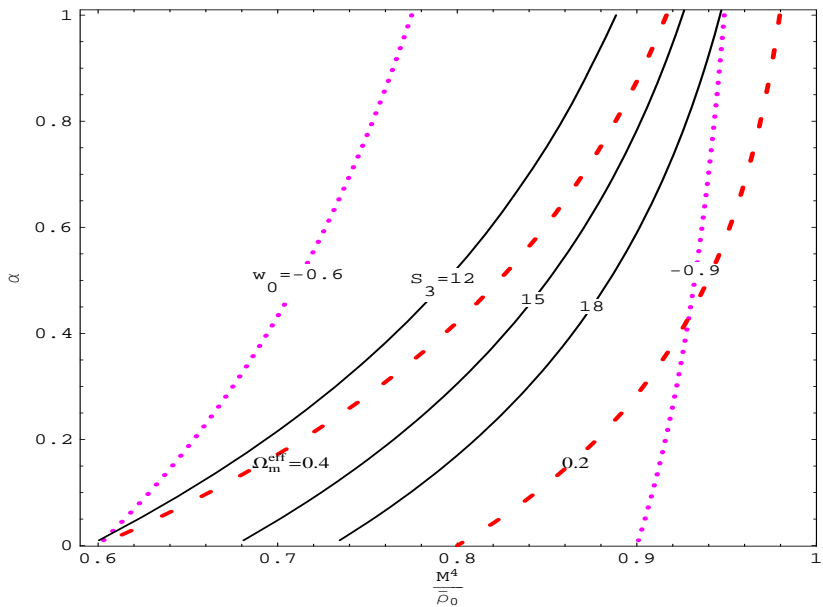}%
\caption{Contours of constant skewness in the
($M^4/\bar{\rho}_0,\alpha$) plane for QCM (solid curves). We also
plot curves of constant $\Omega_m^{eff}$ (dashed curves) and
constant present value of the equation of state $w_0$ (doted
curves).}
\end{figure}

We now consider the following quartessence models:
\begin{eqnarray}
\bar{P}&=&-\frac{M^{4(\alpha+1)}}{\bar{\rho}^{\alpha}}
\,\,\;\;\;\;
\text{(Chaplygin Quartessence)}, \\
\bar{P}&=&-\frac{M^{4}}{\left(\ln\frac{\bar{\rho}}{M^{4}}\right)^{\alpha}}
\,\,\;\;\;\;
\text{(Logarithmic Quartessence)}, \\
\bar{P}&=&-M^{4}\exp{\left(\frac{-\alpha
\bar{\rho}}{M^{4}}\right)} \,\,\;\;\;\; \text{(Exponential
Quartessence)}.
\end{eqnarray}
When $\alpha=0$, all the models above have $\Lambda$CDM as a
limiting case for the background. At first and higher order in
perturbation theory, however, these quartessence models with
$\alpha=0$, have distinct behavior as compared to that of
dark-matter in $\Lambda$CDM.

In Figs. $1$, $2$ and $3$ we show contours of constant skewness
values at present time for the above models. In our numerical
computation, we consider that at $z=10^3$, the growing modes $D_1$
and $D_2$ assume the Einstein-de Sitter behavior: $D_1 \propto a$,
$D_2\propto a^2$, and such that $S_3=34/7$ at that time. We also
include in the figures two curves with constant effective matter
density parameter ($\Omega_m^{eff} =0.2$ and $0.4$). Here,
$\Omega_m^{eff} = \lim_{a\rightarrow 0}\bar{\rho}/\bar{\rho}_0
a^3$, where $\bar{\rho}_0$ is the present value of the background
energy density. We should expect the region between these two
curves to be, roughly, the one allowed by current observational
data. For instance, in \cite{makler03b} it was shown that, for
QCM, constraints from cluster x-ray data (that are the most
restrictive ones), essentially correspond to these curves. The
same holds for the other quartessence models. From the figures it
is clear that, for all the considered models, the skewness in
these regions assumes values between $S_3 \approx 13 - 20$. This
strongly contrasts with what is expected in $\Lambda$CDM and
quintessence, where one obtains $S_3 \approx 5$, weakly sensitive
to the cosmological parameters
\cite{bernardeau94,gaztanaga01,benabed01,multamaki03,multamaki03b}.

At first sight this difference could be interpreted as an
indication that quartessence models are inconsistent with
large-scale skewness measurements \cite{skewobs}. However, care
should be taken when analyzing this issue. Whereas in the
discussion above baryons were neglected, measurements of skewness
from large-scale galaxy distribution, are based on counting
luminous objects, not the dark component. Eqs. (1)-(5) can be
easily generalized to include baryons \cite{makler04}. The main
outcome is that the quartessence skewness is not substantially
affected by the presence of a small amount
($\Omega_{b0}\approx0.04$) of baryons, but on the other hand, the
baryonic skewness ($S_{3b}$) is nearly constant with redshift,
i.e. $S_{3b}\simeq34/7$. Therefore, again baryons behave
differently from the dark component as in the adiabatic
quartessence power-spectrum case \cite{beca03}. We remark that
this holds even for $\alpha =0$.

Further investigation is necessary to clarify to what extent do
skewness measurements from galaxy distribution constrain
quartessence models. Although more challenging to observe, a
potentially powerful probe is the lensing (convergence) skewness
\cite{bernardeau02}, that is sensitive to both baryons and
quartessence. According to \cite{skewlens} current lensing
observations are still too noisy to allow strong constraints on
cosmological parameters. On the other hand, as discussed above,
quartessence predictions for the skewness are quite different from
$\Lambda$CDM models and its variants. Thus, we expect that present
and upcoming data might discriminate among these two classes of
models.

In the present work, we have not quantitatively compared
quartessence predictions with observations. Our goal here is more
modest; we essentially use Figs. $1$, $2$ and $3$ to stress out
the important fact that non-adiabatic quartessence and concordance
models like $\Lambda$CDM and quintessence could be observationally
distinguished by skewness measurements. We leave this
investigation for a future work \cite{makler04}. Finally, we
remark that our results do not apply straightforwardly to
quartessence models with a scalar field with, for instance, a
non-canonical kinetic term \cite{scherrer}, but these models have
yet to be tested against background and power spectrum data.

\begin{figure}\centering \hspace*{-0.8in}
\includegraphics[height= 8 cm,width=11cm]{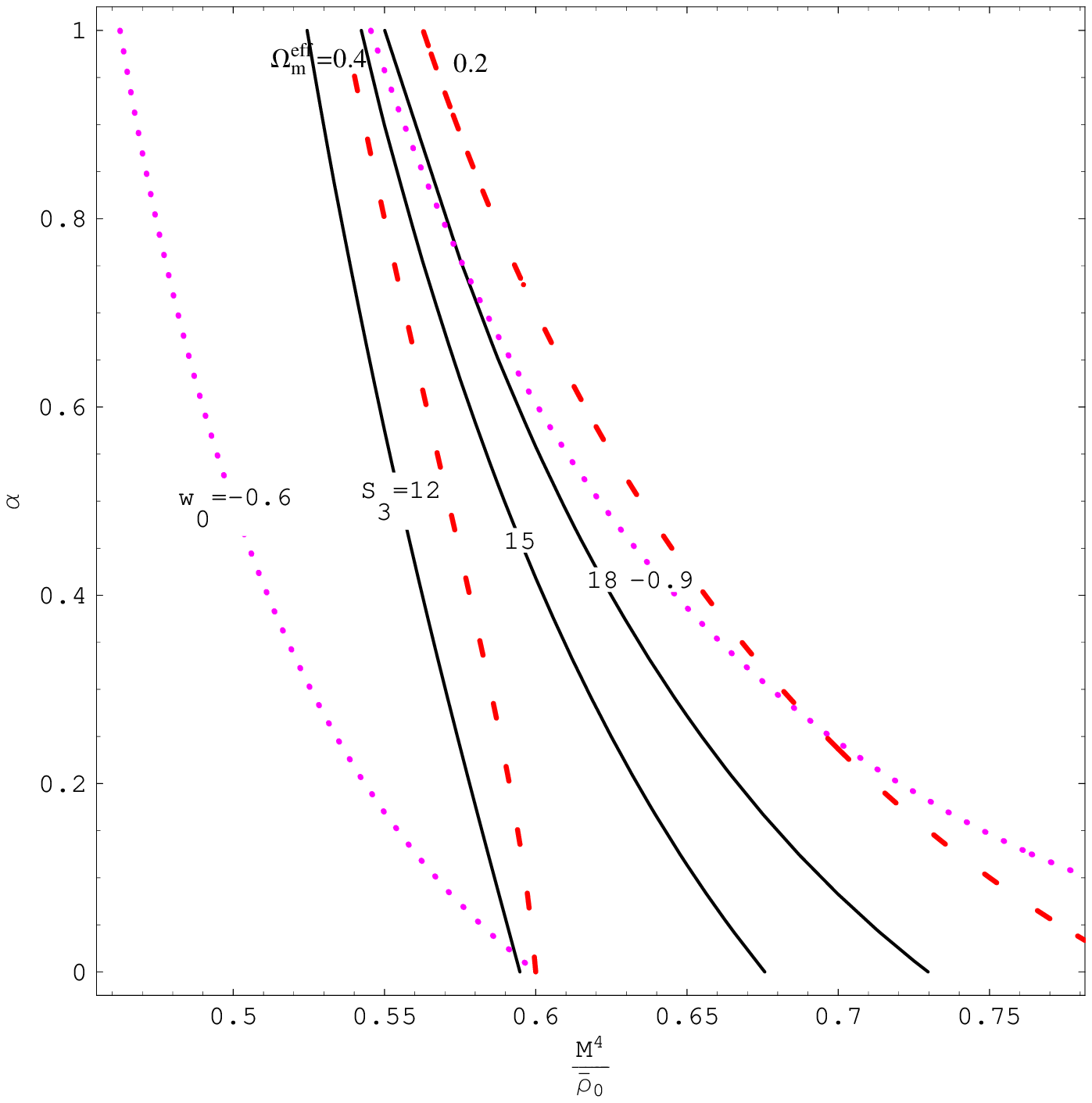}%
\caption{Contours of constant skewness in the
($M^4/\bar{\rho}_0,\alpha$) plane for logarhitmic quartessence
(solid curves). We also plot curves of constant $\Omega_m^{eff}$
(dashed curves) and constant present value of the equation of
state $w_0$ (doted curves). }
\end{figure}

\begin{figure}\centering \hspace*{-0.8in}
\includegraphics[height= 8 cm,width=11.0cm]{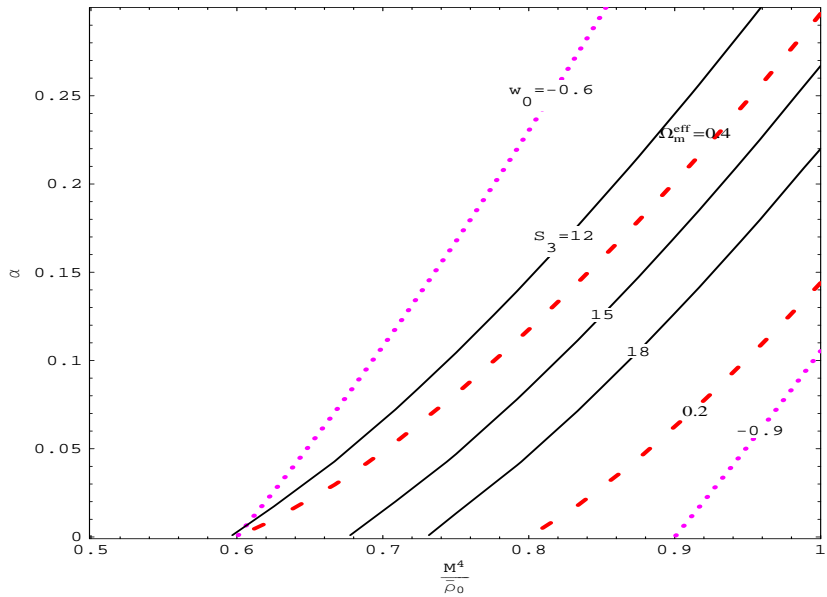}%
\caption{Contours of constant skewness in the
($M^4/\bar{\rho}_0,\alpha$) plane for exponential quartessence
(solid curves). We also plot curves of constant $\Omega_m^{eff}$
(dashed curves) and constant present value of the equation of
state $w_0$ (doted curves). }
\end{figure}

\acknowledgments

The authors would like to thank the Brazilian research agencies
CAPES, CNPq and FAPERJ for financial support.


\begin{thebibliography}{99}

\bibitem{makler03} M. Makler, S.Q. Oliveira, and I. Waga, Phys. Lett.
B \textbf {555}, 1 (2003).

\bibitem{kamenshchik01} A. Kamenshchik, U. Moschella, and V. Pasquier, Phys.
Lett. B \textbf {511}, 265 (2001); M. Makler, {\it Gravitational
Dynamics of Structure Formation in the Universe}, PhD Thesis,
Brazilian Center for Research in Physics (2001); N. Bili\'{c},
G.B. Tupper, and R.D. Viollier, Phys. Lett. B \textbf {535}, 17
(2002); M.C. Bento, O. Bertolami, and A.A. Sen, Phys. Rev. D
\textbf {66}, 043507 (2002).

\bibitem{makler03b} M. Makler, S.Q. Oliveira, and I. Waga, Phys. Rev. D
\textbf{68}, 123521 (2003).

\bibitem{amendola03} L. Amendola, F. Finelli, C. Burigana, and D.
Carturan, JCAP \textbf{07}, 005 (2003). See also D. Carturan, F.
Finelli, Phys.Rev. D \textbf{68}, 103501 (2003) and R. Bean and O.
Dore, Phys.Rev. D \textbf{68}, 023515 (2003) for the case of the
Chaplygin fluid as only dark energy.

\bibitem{beca03} L.M.G. Be\c{c}a, P.P. Avelino, J.P.M. de Carvalho, and C.J.A.P.
Martins, Phys. Rev. D \textbf{67}, 101301(R) (2003).

\bibitem{sandvik02} H. Sandvik, M. Tegmark, M. Zaldarriaga, and I. Waga,
\texttt{astro-ph/0212114} (v2), to appear Phys. Rev. D (2004).

\bibitem{reis03b} R.R.R. Reis, I. Waga, M.O. Calv\~{a}o, and S.E. Jor\'{a}s,
Phys. Rev. D \textbf{68},  061302(R) (2003).

\bibitem{avelino03} P.P. Avelino, L.M.G. Be\c{c}a, J.P.M. de Carvalho,
C.J.A.P. Martins, and E.J. Copeland,  Phys. Rev. D \textbf{69},
041301(R) (2004).

\bibitem{obs} Since $\delta P =0$, for non-adiabatic quartessence models we always have $<P>/\bar{P}
=1$.

\bibitem{bernardeau92} F. Bernardeau, Astrophys. J. \textbf{392}, 1
(1992).

\bibitem{bernardeau94} F. Bernardeau, Astrophys. J. \textbf{433}, 1
(1994).

\bibitem{fosalba98} P. Fosalba and E. Gazta\~{n}aga,  Mon. Not. R. Astron. Soc. \textbf{301}, 503
(1998).

\bibitem{lima} J.A.S. Lima, V. Zanchin, and R. Brandemberger, Mon. Not. R. Astron. Soc. \textbf{291},
L1 (1997).

\bibitem{reis03} R.R.R. Reis, Phys. Rev. D \textbf{
67}, 087301 (2003), Erratum-ibid. D \textbf{68}, 089901 (2003).

\bibitem{ellis} G.F.R. Ellis, Relativistic Cosmology. In R. K. Sachs,
editor, General Relativity and Cosmology, proceedings of the XLVII
Enrico Fermi Summer School, Academic Press, 1971.

\bibitem{hu} W. Hu, Astrophys.J. \textbf{506}, 485 (1998).

\bibitem{peebles} P.J.E. Peebles, The Large Structure of the Universe, Princeton University Press, Princeton, (1980).

\bibitem{fry} J. Fry, Astrophys.J. \textbf{279}, 499 (1984).

\bibitem{gaztanaga01} E. Gazta\~{n}aga and J.A. Lobo,  Astrophys. J. \textbf{548},
47 (2001).

\bibitem{benabed01} K. Benabed and F. Bernardeau, Phys. Rev.
D \textbf{64}, 083501 (2001).

\bibitem{multamaki03} T. Multam\"{a}ki, E. Gazta\~{n}aga, and M. Manera,  Mon. Not. R. Astron. Soc. \textbf{344},
761 (2003).

\bibitem{multamaki03b} T. Multam\"{a}ki, M. Manera, and E. Gazta\~{n}aga, Phys. Rev. D \textbf{69}, 023004 (2004).

\bibitem{makler04} M. Makler {\it et al.}, in preparation (2004).

\bibitem{skewobs} E. Gazta\~{n}aga and J.A. Frieman,  Astrophys. J. \textbf{437}, L13 (1994); F. Hoyle, I.
Szapudi, and C.M. Baugh,  Mon. Not. R. Astron. Soc. \textbf{317},
L51 (2000); I. Szapudi, M. Postman, T.R. Lauer, and W. Oegerle,
Astrophys. J. \textbf{548}, 114 (2001); I. Szapudi {\it et. al.},
Astrophys. J. \textbf{570}, 75 (2002); for review see F.
Bernardeau, S. Colombi, E. Gazta\~{n}aga, and R. Scoccimarro,
Phys. Rep. \textbf{367}, 1 (2002).

\bibitem{bernardeau02} F. Bernardeau, L. van Waerbeke , and Y. Mellier,
Astron. Astrophys. \textbf{322}, 1 (1997).

\bibitem{skewlens} F. Bernardeau, Y. Mellier, and L. van Waerbeke, Astron. Astrophys. \textbf{389}, L28 (2002); U.-L. Pen {\it et al.}, Astrophys. J. \textbf{592}, 664
(2003).

\bibitem{scherrer} R. J. Scherrer, astro-ph/0402316 (v2).

\end{thebibliography}
\end{document}